\documentclass[twoside]{ilcws08}
\usepackage[latin1]{inputenc}
\usepackage[dvips]{graphicx,epsfig,color}
\usepackage{wrapfig,rotating}
\usepackage{amssymb,amsmath,array}

\begin{document}
\title{
Higgs boson pair production at the Photon Linear Collider in the two Higgs doublet model} 
\author{Eri~Asakawa$^1$, Daisuke~Harada$^{2,3}$, Shinya~Kanemura$^4$, Yasuhiro~Okada$^{2,3}$ and Koji~Tsumura$^5$
\vspace{.3cm}\\
1- Institute of Physics, Meiji Gakuin University \\
Yokohama 244-8539, Japan\\
\vspace{.1cm}\\
2- KEK Theory Center, Institute of Particle and Nuclear Studies, KEK \\
1-1 Oho, Tsukuba, Ibaraki 305-0801, Japan\\
\vspace{.1cm}\\
3- Department of Particle and Nuclear Physics,\\
the Graduate University for Advanced Studies (Sokendai) \\
1-1 Oho, Tsukuba, Ibaraki 305-0801, Japan\\
\vspace{.1cm}\\
4- Department of Physics, University of Toyama \\
3190 Gofuku, Toyama 930-8555, Japan\\
\vspace{.1cm}\\
5- International Centre for Theoretical Physics \\
Strada Costiera 11, 34014 Trieste, Italy\\
}

\maketitle

\begin{abstract}
We calculate the cross section of the lightest Higgs boson pair production at the Photon Linear Collider in the two Higgs doublet model. 
We focus on the scenario in which the lightest Higgs boson has the standard model like couplings to gauge bosons.
We take into account the one-loop correction to the $hhh$ coupling as well as additional one-loop diagrams due to charged bosons to the $\gamma\gamma \to hh$ helicity amplitudes.
We discuss the impact of these corrections on the $hhh$ coupling measurement at the Photon Linear Collider.
\end{abstract}

\section{Introduction}
The Higgs sector is the last unknown part of the standard model (SM).
In the SM, the tree level Higgs self-coupling $\lambda_{hhh} = 3m_{h}^{2}/v$ and $\lambda_{hhhh} = 3m_{h}^{2}/v^{2}$ are uniquely determined by the Higgs boson mass $m_{h}$, where $v$ is vacuum expectation value (VEV) of the Higgs boson.
The effective Higgs potential is written as
\begin{eqnarray}
V = \frac{1}{2} m_{h}^{2} h^{2} + \frac{1}{3!} \tilde{\lambda}_{hhh} h^{3} + \frac{1}{4!} \tilde{\lambda}_{hhhh} h^{4} + \cdots,
\end{eqnarray}
where the effective Higgs self-couplings $\tilde{\lambda}_{hhh}$ and $\tilde{\lambda}_{hhhh}$ are given by precision measurement of $hhh$ and $hhhh$ couplings.
If the deviation from the SM tree level Higgs self-coupling ($\lambda_{hhh}$ and $\lambda_{hhhh}$) is found, it can be regarded as an evidence of new physics beyond the SM.
The origin of the spontaneous electroweak symmetry breaking (EWSB) would be experimentally tested after the discovery of a new scalar particle by measuring its mass and self-couplings.
The Higgs self-coupling measurement is one of main purposes at the International Linear Collider (ILC).
The structure of the Higgs potential depends on the scenario of new physics beyond the SM, so that precision measurement of the $hhh$ coupling can be a probe of each new physics scenario\cite{hhh-thdm1,hhh-thdm2}.

It is known that the measurement of the triple Higgs boson coupling is rather challenging at the CERN Large Hadron Collider (LHC).
At the SLHC with luminosity of 3000 ${\rm fb}^{-1}$, the $hhh$ coupling can be determined with an accuracy of 20-30$\%$ for 160 GeV $\leq m_h \leq$ 180 GeV\cite{hhh-lhc,hhh-lhc-ilc}.
At the ILC, the main processes for the $hhh$ measurement are the double Higgs boson production mechanisms via the Higgs-strahlung and the W-boson fusion\cite{eehhZ1,eehhZ2}.
At the ILC with a center of mass energy of 500 GeV, the double Higgs strahlung process $e^+e^- \to Zhh$ is dominant.
On the other hand, W-boson fusion process $e^+e^- \to hh\nu\bar{\nu}$ becomes dominant due to its $t$-channel nature at 1 TeV or higher energies\cite{eehhZ3}.
Sensitivity to the $hhh$ coupling in these processes becomes rapidly worse for greater Higgs boson masses.
In particular, for the intermediate mass range (140 GeV $\leq m_h \leq$ 200 GeV), it has not yet been known how accurately the $hhh$ coupling can be measured by the electron-positron collision.
The Photon Linear Collider (PLC) is an optional experiment of the ILC.
The possibility of measuring the $hhh$ coupling via the process of $\gamma \gamma \to hh$ has been discussed in Ref.~\cite{Jikia}.
In Ref.~\cite{Belusevic} the statistical sensitivity to the $hhh$ coupling constant has been studied especially for a light Higgs boson mass in relatively low energy collisions.

In this paper, we study the double Higgs production process at the PLC.
In Sect.~\ref{sect2}, we discuss the statistical sensitivity to the $hhh$ coupling constant via the process of $e^- e^- \to \gamma \gamma \to hh$ at the PLC in the SM.
In Sect.~\ref{sect3}, we study the new particle effects on the $\gamma \gamma \to hh$ process in the two Higgs doublet model (THDM).

\section{The statistical sensitivity to the $hhh$ coupling constant}\label{sect2}
We study the statistical sensitivity to the $hhh$ coupling constant for wide regions of the Higgs boson masses and the collider energies at the PLC.
The $\gamma \gamma \to hh$ process is an one-loop induced process.
The Feynman diagrams for this process in the SM are given in Ref.~\cite{Jikia}.
There are two types of diagrams, which are the pole diagrams and the box diagrams.
The amplitude of the pole diagrams describes as ${\cal M}_{{\rm pole}} \propto \tilde{\lambda}_{hhh}/s$, where $\sqrt{s}$ is the center of mass energy of the $\gamma \gamma$ system.
It is suppressed by $1/s$ at the high energy region, so that the statistical sensitivity to the $hhh$ coupling becomes rapidly worse for this region.
On the other hand, the box diagrams do not depend on the $hhh$ coupling.
\begin{figure}[t]
\begin{center}
\includegraphics[width=6.5cm]{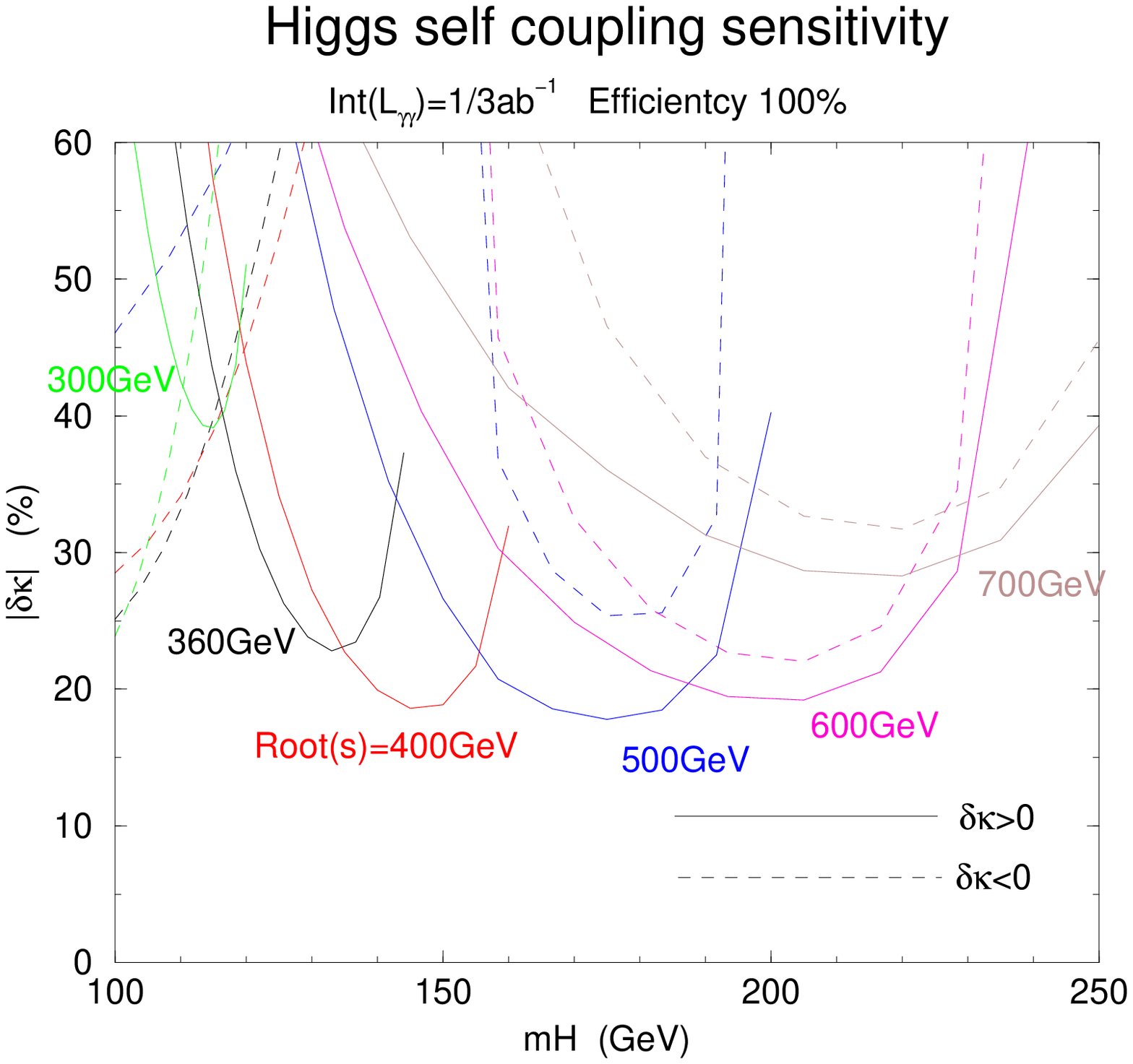}
\includegraphics[width=6.5cm]{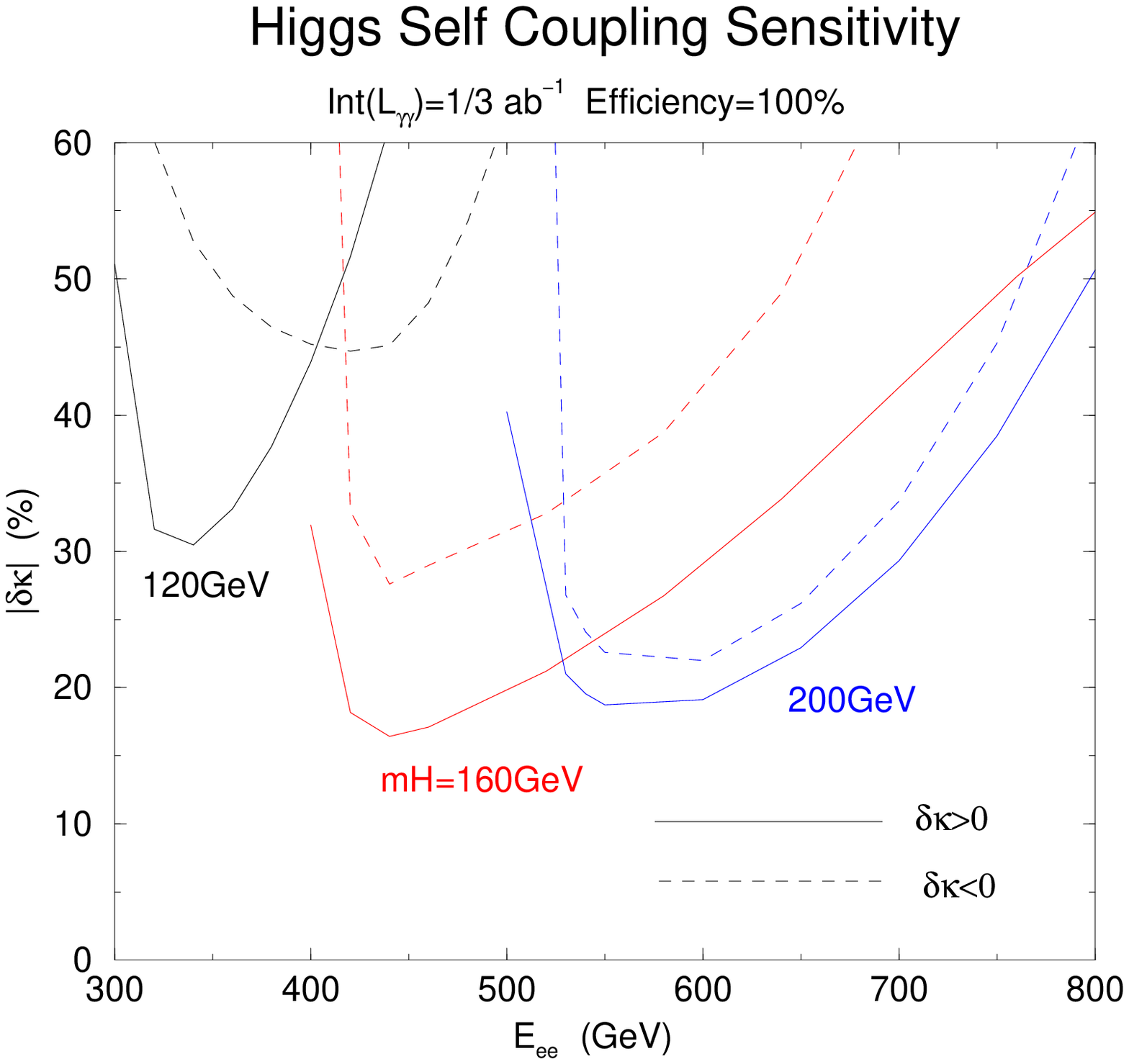}
\caption{
The statistical sensitivity to the $hhh$ coupling constant at the PLC.
In the left [right] figure, the statistical sensitivity is shown as a function of $m_h$ [$E_{ee}$] for each value of $E_{ee}$ [$m_h$].
Solid [Dotted] lines correspond to $\delta \kappa > 0$ [$\delta \kappa < 0$] case.
}
\label{figure1}
\end{center}
\end{figure}

In Fig.~\ref{figure1}, we present the statistical sensitivity on the Higgs self-coupling constant at the PLC.
We modify the triple Higgs coupling constant as $\tilde{\lambda}_{hhh} = \lambda_{hhh}(1 + \delta \kappa)$, where $\delta \kappa$ represents deviation from the SM prediction.
We assume that the efficiency of the particle tagging is 100$\%$ with an integrated luminosity of $1/3$ ${\rm ab}^{-1}$ and $E_{ee}$ is the center of mass energy of the $e^- e^-$ system.
We plot $\delta \kappa$ based on statistical error of the event number in the $e^- e^- \to \gamma \gamma \to hh$ process in the SM.
Namely, $\delta \kappa$ is determined by
\begin{eqnarray}
\left| N(\delta \kappa) - N(\delta \kappa = 0) \right| = \sqrt{N(\delta \kappa =0)},
\end{eqnarray}
for assumed luminosity.
Notice that $\delta \kappa$ is not symmetric with respect to $\delta \kappa =0$ because there is interference between pole and box diagrams.
The cases for $\delta \kappa > 0$ and $\delta \kappa < 0$ are shown separetly.
The left [right] figure shows the sensitivity as a function of $m_{h}$ [$E_{ee}$].
It is found that when the collision energy is limited to be lower than 500-600 GeV the statistical sensitivity to the $hhh$ coupling can be better for the process in the $\gamma \gamma$ collision than that in the electron-positron collision for the Higgs boson with the mass of 160 GeV\cite{tilc08}.
%

\section{The $\gamma \gamma \to hh$ process in the THDM}\label{sect3}
We consider the new particle effects on the $\gamma \gamma \to hh$ process in the THDM, in which additional CP-even, CP-odd and charged Higgs boson appear.
It is known that non-decoupling loop effect of extra Higgs bosons shift the $hhh$ coupling value from the SM by ${\cal O}(100)\%$\cite{hhh-thdm1}.
In the $\gamma \gamma \to hh$ helicity amplitudes, there are additional one-loop diagrams by the charged Higgs boson loop to the ordinary SM diagrams (the W-boson loop and the top quark loop).
It is found that both the charged Higgs boson loop contribution to the $\gamma \gamma \to hh$ amplitudes and the non-decoupling effect on the $hhh$ coupling can enhance the cross section from its SM value significantly\cite{gamgamHH-thdm}.

In order to study the new physics effect on $\gamma \gamma \to hh$ process, we calculate the helicity amplitudes in the THDM.
The THDM Higgs potential is given by
\begin{eqnarray}
 V_{\rm THDM}&=& \mu_1^2 |\Phi_1|^2+\mu_2^2 |\Phi_2|^2-(\mu_3^2
  \Phi_1^\dagger \Phi_2 + {\rm h.c.})\nonumber\\
&&  + \lambda_1 |\Phi_1|^4
  + \lambda_2 |\Phi_2|^4
 + \lambda_3 |\Phi_1|^2|\Phi_2|^2
  + \lambda_4 |\Phi_1^\dagger \Phi_2|^2
  + \frac{\lambda_5}{2} \left\{(\Phi_1^\dagger \Phi_2)^2 + {\rm h.c.}
                        \right\}, 
\end{eqnarray}
where $\Phi_{1}$ and $\Phi_{2}$ are two Higgs doublets with hypercharge $+1/2$.
The Higgs doublets are parametrized as
\begin{eqnarray}
  \Phi_i = \left[ \begin{array}{c}
            \omega_i^+ \\ \frac{1}{\sqrt{2}}(v_i+h_i + i z_i)
            \end{array}
   \right], \hspace{4mm} (i=1,2),  
\end{eqnarray}
where VEVs $v_1$ and $v_2$ satisfy $v_1^2+v_2^2 = v^2 \simeq (246 \hspace{2mm} {\rm GeV})^2$.
The mass matrices can be diagonalized by introducing the mixing angles $\alpha$ and $\beta$, where $\alpha$ diagonalizes the mass matrix of the CP-even neutral bosons, and $\tan\beta=v_2/v_1$.
Consequently, we have two CP-even ($h$ and $H$), a CP-odd ($A$) and a pair of charged ($H^\pm$) bosons.
We define $\alpha$ such that $h$ is the SM-like Higgs boson when $\sin(\beta-\alpha)=1$.

We concentrate on the case with so called the SM-like limit [$\sin(\beta - \alpha) = 1$], where the lightest Higgs boson $h$ has the same tree-level couplings as the SM Higgs boson, and the other bosons do not couple to gauge bosons and behave just as extra scalar bosons.
In this limit, the masses of Higgs bosons are
\begin{eqnarray}
    m_h^2 &=& \{\lambda_1 \cos^4\beta+\lambda_2 \sin^4\beta + 2
     (\lambda_3+\lambda_4+\lambda_5) \cos^2\beta\sin^2\beta\} v^2,\\
    m_H^2 &=& M^2 +
     \frac{1}{8}\left\{\lambda_1+\lambda_2-2(\lambda_3+\lambda_4+\lambda_5)\right\}(1-\cos
      4\beta)v^2,\\
    m_A^2 &=& M^2 - \lambda_5 v^2,\\
    m_{H^\pm}^2 &=& M^2 - \frac{\lambda_4+\lambda_5}{2}v^2, 
\end{eqnarray}
where $M (= |\mu_3|/\sqrt{\sin\beta\cos\beta})$ represents the soft breaking scale for the discrete symmetry, and determines the decoupling property of the extra Higgs bosons.
When $M \sim 0$, the extra Higgs bosons $H$, $A$ and $H^\pm$ receive their masses from the VEV, so that the masses are proportional to $\lambda_i$.
Large masses cause significant non-decoupling effect in the radiative correction to the $hhh$ coupling.
On the other hand, when $M \gg v$ the masses are determined by $M$.
In this case, the quantum effect decouples for $M \to \infty$.

It is known that in the THDM $\lambda_{hhh}$ can be changed from the SM prediction by the one-loop contribution of extra Higgs bosons due to the non-decoupling effect (when $M \sim 0$).
In the following analysis, we include such an effect on the cross sections.
The effective $hhh$ coupling $\Gamma_{hhh}^{\rm THDM}(\hat{s},m_h^2,m_h^2)$ is evaluated at the one-loop level as\cite{hhh-thdm1}
\begin{eqnarray}
\Gamma_{hhh}^{\rm THDM}(\hat{s},m_h^2,m_h^2) \simeq  \frac{3 m_h^2}{v} \left[ 1 + \sum_{\Phi=H,A,H^+,H^-}
                               \frac{m_{\Phi}^4}{12 \pi^2 v^2 m_h^2}
                              \left(1-\frac{M^2}{m_\Phi^2}\right)^3 -
                              \frac{N_c m_t^4}{3\pi^2 v^2 m_h^2} \right]. \label{hhh-2hdm}
\end{eqnarray}
The exact one-loop formula for $\Gamma_{hhh}^{\rm THDM}$ is given in
Ref.~\cite{hhh-thdm2}, which has been used in our actual numerical analysis.

\begin{figure}[t]
\begin{center}
\includegraphics[width=6.5cm]{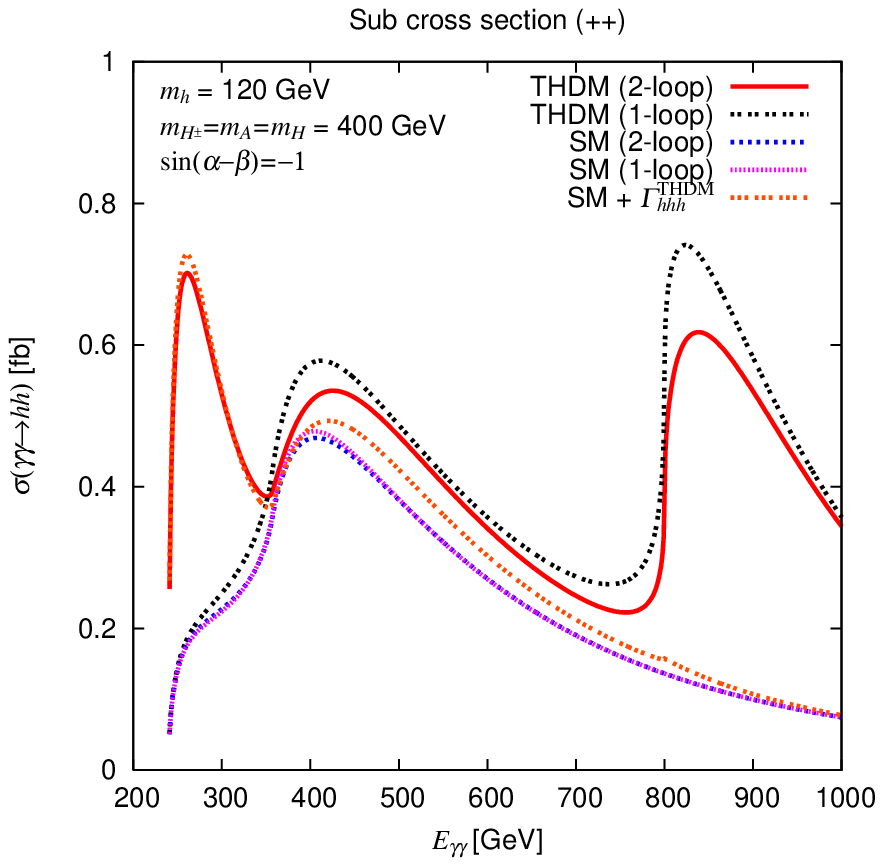}
\includegraphics[width=6.5cm]{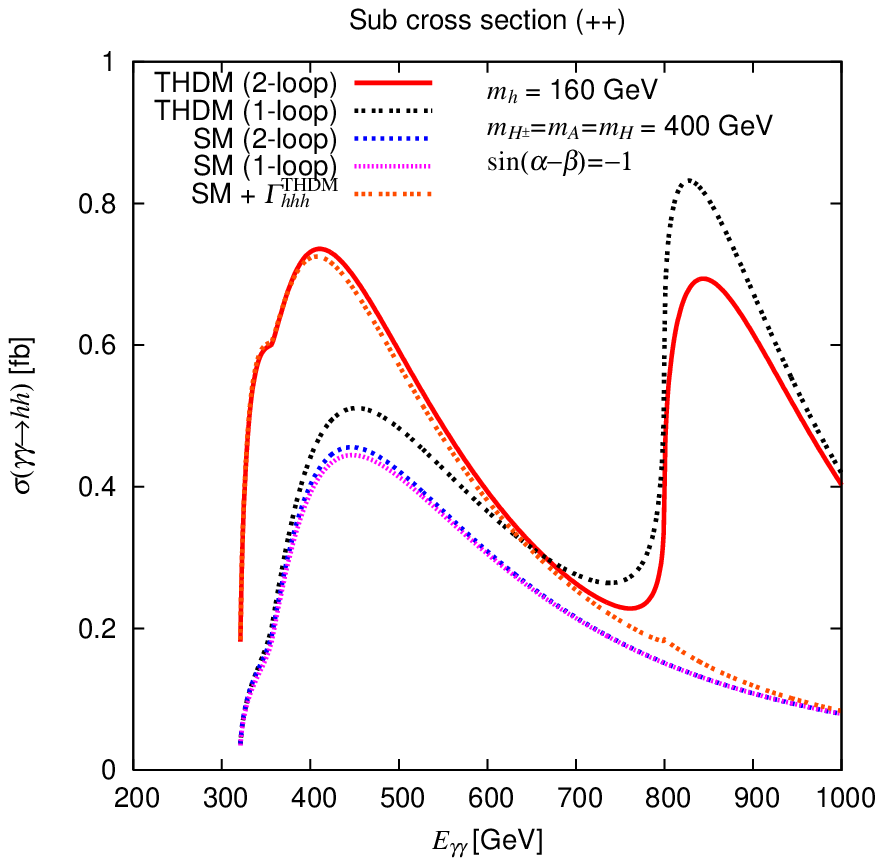}
\caption{
The cross section $\hat{\sigma}(+,+)$ for the sub process $\gamma \gamma \to hh$ with the photon helicity set $(+,+)$ as a function of the collision energy $E_{\gamma \gamma}$. In the left [right] figure the parameters are taken to be $m_{h} = 120$ [$160$] GeV for $m_{\Phi}(\equiv m_{H} = m_{A} = m_{H^{\pm}}) = 400$ GeV, $\sin(\beta - \alpha) = 1$, $\tan \beta =1$ and $M=0$.
}
\label{figure2}
\end{center}
\end{figure}
In Fig.~\ref{figure2}, we plot the cross sections of $\gamma \gamma \to hh$ for the helicity set $(+,+)$ as a function of the photon-photon collision energy $E_{\gamma \gamma}$.
The five curves correspond to the following cases,
\begin{itemize}
\item[(a)]
     THDM 2-loop: the cross section in the THDM with additional one-loop corrections to the
$hhh$ vertex, $\Gamma_{hhh}^{\rm THDM}$.

\item[(b)]
 THDM 1-loop: the cross section  in the THDM with the tree level $hhh$ coupling constant $\lambda_{hhh}$.

\item[(c)]
SM 2-loop: the cross section in the SM with additional top loop correction to the $hhh$
coupling $\Gamma_{hhh}^{\rm SM}$ given in Ref.~\cite{hhh-thdm2}.

\item[(d)]
SM 1-loop: the cross section in the SM with the
tree level $hhh$ coupling constant $\lambda_{hhh}^{\rm SM}$ ($=\lambda_{hhh}$ for $\sin(\beta-\alpha)=1$).  

\item[(e)]
 For comparison, we also show the result which corresponds to
the SM 1-loop result with the effective $hhh$ coupling 
$\Gamma^{\rm THDM}_{hhh}$.
\end{itemize}
In the left figure, there are three peaks in the case (a) (THDM 2-loop).
The one at the lowest $E_{\gamma \gamma}$ is the peak just above the threshold of $hh$ production.
There the cross section is by about factor three enhanced as compared to the SM prediction due to the effect of $\Delta \Gamma_{hhh}^{\rm THDM}/\Gamma_{hhh}^{\rm SM}$ ($\sim 120\%$) because of the dominance of the pole diagrams in $\gamma \gamma \to hh$.
The second peak at around $E_{\gamma \gamma} \sim 400$ GeV comes from the top quark loop contribution which is enhanced by the threshold of top pair production. 
Around this point, the case (a) can be described by the case (e) (SM+$\Gamma_{hhh}^{\rm THDM}$).
For $E_{\gamma \gamma} \sim 400$-$600$ GeV, the cross section in the case (a) deviates from the case (c) (SM 2-loop) due to both the charged Higgs loop effect and the effect of $\Delta \Gamma_{hhh}^{\rm THDM}/\Gamma_{hhh}^{\rm SM}$.
The third peak at around $E_{\gamma \gamma} \sim 850$ GeV is the threshold enhancement of the charged Higgs boson loop effect, where the real production of charged Higgs bosons occurs.
The contribution from the non-pole one-loop diagrams are dominant.
In the right figure, we can see two peaks around $E_{\gamma \gamma} \sim 350$-$400$ GeV and $850$ GeV.
At the first peak, the contribution from the pole diagrams is dominant so that the cross section is largely enhanced by the effect of $\Delta \Gamma_{hhh}^{\rm THDM}/\Gamma_{hhh}^{\rm SM}$ by several times 100\% for $E_{\gamma\gamma} \sim 350$ GeV.
It also amounts to about 80\% for $E_{\gamma\gamma} \sim 400$ GeV.
For $E_{\gamma\gamma} < 600$-700 GeV, the result in the case (e) gives a good description of that in the case (a).
The second peak is due to the threshold effect of the real $H^+H^-$ production as in the left figure.

\begin{figure}[t]
\begin{center}
\includegraphics[width=6.5cm]{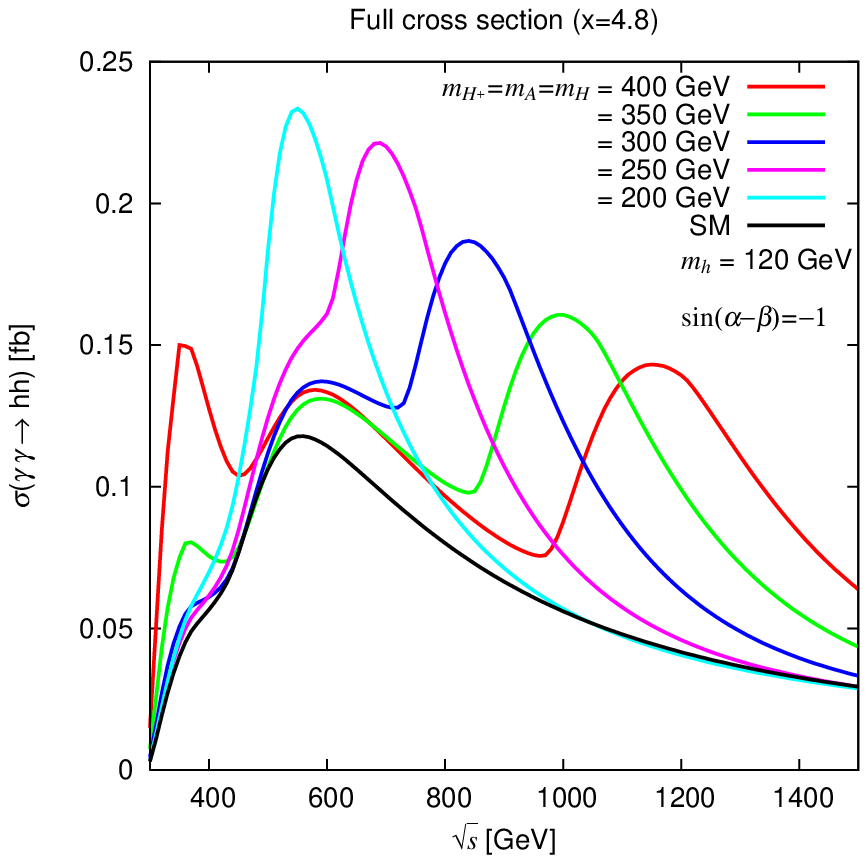}
\includegraphics[width=6.5cm]{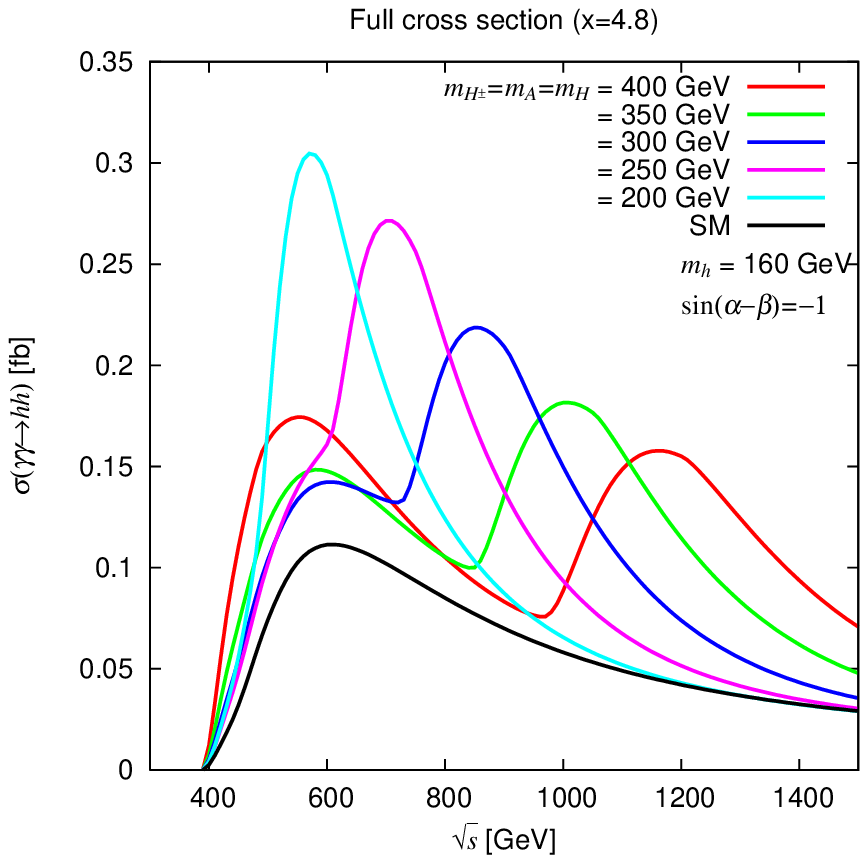}
\caption{
The full cross section of $e^{-}e^{-} \to \gamma \gamma \to hh$ as a function of $\sqrt{s}$ for each value of $m_{\Phi}(=m_{H} = m_{A} = m_{H^{\pm}})$ with $\sin(\beta - \alpha) = 1$, $\tan\beta = 1$ and $M=0$. The case for $m_{h} = 120$ [$160$] GeV is shown in the left [right] figure.
}
\label{figure3}
\end{center}
\end{figure}
In Fig.~\ref{figure3}, the full cross section of $e^- e^- \to \gamma \gamma \to hh$ is given from the sub cross sections by convoluting the photon luminosity spectrum\cite{Jikia}.
In our study, we set $x = 4E_{b}\omega_{0}/m_{e}^{2} = 4.8$ where $E_{b}$ is the energy of electron beam, $\omega_{0}$ is the laser photon energy and $m_{e}$ is the electron mass.
In order to extract the contribution from $\hat{\sigma}(+,+)$ that is sensitive to the $hhh$ vertex, we take the polarizations of the initial laser beam to be both $-1$, and those for the initial electrons to be both $+0.45$.
The full cross section for $m_{\Phi} = 400$ GeV has similar energy dependences to the sub cross section $\hat{\sigma}(+,+)$ in Fig.~\ref{figure2}, where corresponding energies are rescaled approximately by around $\sqrt{s} \sim E_{\gamma\gamma}/0.8$ due to the photon luminosity spectrum.
For smaller $m_{\Phi}$, the peak around $\sqrt{s} \sim 350$ GeV becomes lower because of smaller $\Delta \Gamma_{hhh}^{\rm THDM}/\Gamma_{hhh}^{\rm SM}$.

\begin{figure}[t]
\begin{center}
\includegraphics[width=6.5cm]{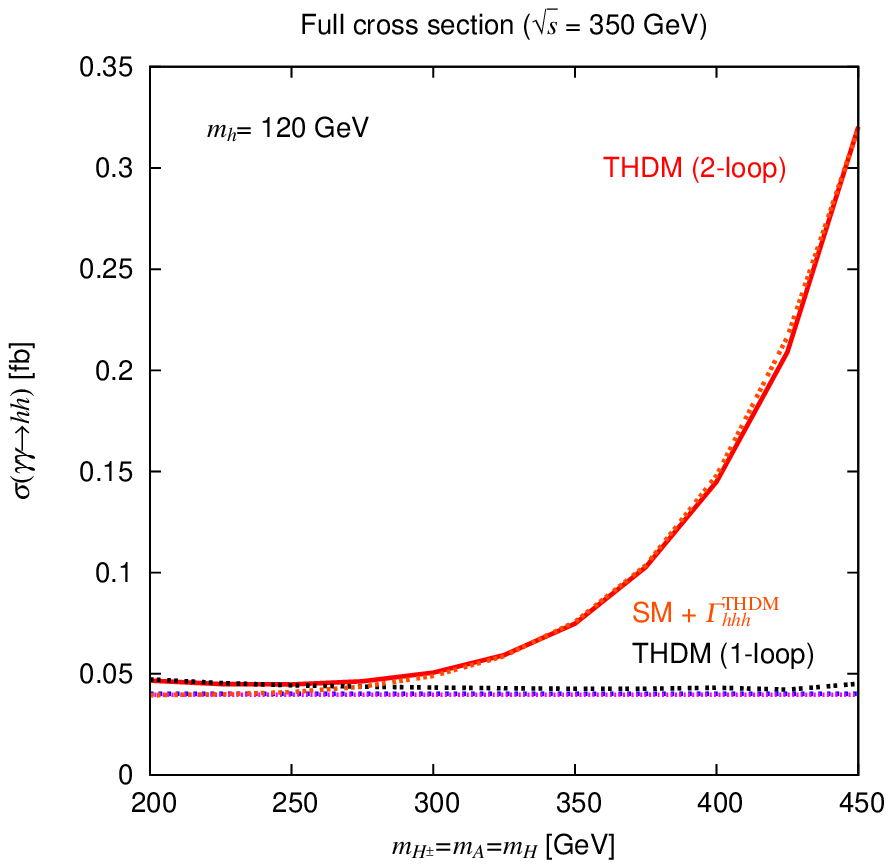}
\includegraphics[width=6.5cm]{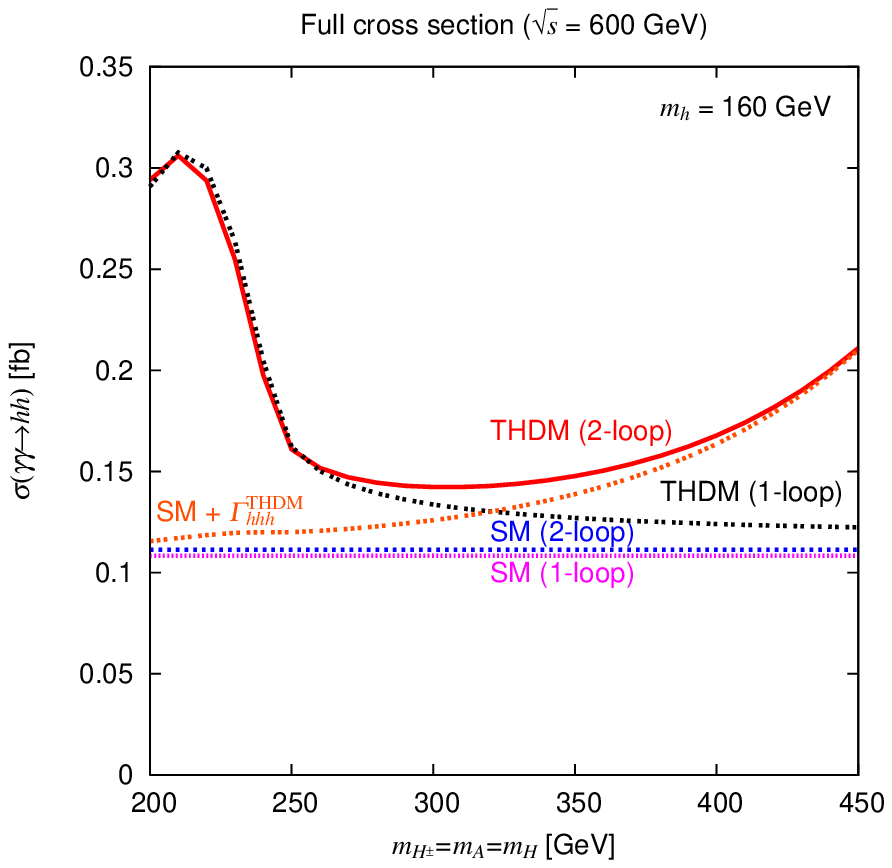}
\caption{
In the left [right] figure, the full cross section of $e^{-}e^{-} \to \gamma \gamma \to hh$ at $\sqrt{s} = 350$ GeV [$600$ GeV] for $m_{h} = 120$ [$160$] GeV is shown as a function of $m_{\Phi}(= m_{H} = m_{A} = m_{H^{\pm}})$ with $\sin(\beta - \alpha) =1$, $\tan \beta = 1$ and $M = 0$.
}
\label{figure4}
\end{center}
\end{figure}
In Fig.~\ref{figure4}, five curves correspond to the cases (a) to (e) in Fig.~\ref{figure2}.
In the left figure, one can see that the cross section is enhanced due to the enlarged $\Gamma_{hhh}^{\rm THDM}$ for larger values of $m_{\Phi}^{}$ which is proportional to $m_{\Phi}^4$ (when $M \sim 0$).
This implies that the cross section for these parameters is essentially determined by the pole diagram contributions.
The effect of the charged Higgs boson loop is relatively small since the threshold of charged Higgs boson production is far.
Therefore, the deviation in the cross section from the SM value is smaller for relatively small $m_{\Phi}^{}$ (10-20\% for $m_{\Phi} < 300$ GeV due to the charged Higgs loop effect) but it becomes rapidly enhanced for greater values of $m_{\Phi}^{}$ (${\cal O}(100)$ \% for $m_{\Phi} > 350$ GeV due to the large $\Delta \Gamma_{hhh}^{\rm THDM}$).
A similar enhancement for the large $m_{\Phi}$ values can be seen in the right figure.
The enhancement in the cross section in the THDM can also be seen for $m_\Phi^{} < 250$ GeV, where the threshold effect of the charged Higgs boson loop appears around $\sqrt{s} \sim 600$ GeV in addition to that of the top quark loop diagrams.
For $m_\Phi^{}=250$-$400$ GeV, both contributions from the charged Higgs boson loop contribution and the effective $hhh$ coupling are important and enhance the cross section from its SM value by 40-50\%.
%

\section{Conclusions}
In this paper, we have analysed the new physics loop effects on the cross section of $\gamma \gamma \to hh$ in the THDM with SM-like limit including the next to leading effect due to the extra Higgs boson loop diagram in the $hhh$ vertex.
Our analysis shows that the cross section can be largely changed from the SM prediction by the two kinds of contributions; i.e., additonal contribution by the charged Higgs boson loop effect, and the effective one-loop $hhh$ vertex $\Gamma_{hhh}^{\rm THDM}$ enhanced by the non-decoupling effect of extra Higgs bosons.
The cross section strongly depends on $m_{h}$ and $\sqrt{s}$ and also on $m_{\Phi}$.
The approximation of the full cross section in the case (a) (THDM 2-loop) by using the result in the case (e) (SM+$\Gamma_{hhh}^{\rm THDM}$) is a good description for  $\sqrt{s} \ll 2m_\Phi/0.8$.
On the other hand, in a wide region between threshold of top pair production and that of charged Higgs boson pair production, both the contributions (those from charged Higgs boson loop effect and from $\Gamma_{hhh}^{\rm THDM}$) are important. 
In the region below  the threshold of the real production of extra Higgs bosons, cross section is largely enhanced from the SM value by the effects of the charged Higgs boson loop and the effective $\Gamma_{hhh}^{\rm THDM}$ coupling.
These New Physics effects would be detectable at the future Photon Linear Collider.
%

\section{Acknowledgments}
The authors would like to thank all the members of the ILC physics subgroup
\cite{softg} for useful discussions. 
This study is supported in part by the Creative Scientific Research Grant
No. 18GS0202 of the Japan Society for Promotion of Science.
The work of S. K. was supported in part by Grant-in-Aid for Science
Research, Japan Society for the Promotion of Science (JSPS),
No. 18034004. The work of Y. O. was supported in part by Grant-in-Aid for 
Science Research, MEXT-Japan, No. 16081211, and JSPS, No. 20244037.
%


\begin{footnotesize}



%

\end{footnotesize}


\end{document}